\documentclass[aps,prl,superscriptaddress,twocolumn,longbibliography]{revtex4-1}
\usepackage{bbm}
\usepackage{graphicx}
\usepackage{dcolumn}
\usepackage{bm}
\usepackage{subfigure}
\usepackage{amsmath}
\usepackage{feynmf}
\usepackage{hyperref}

\usepackage{attachfile}

\newcommand{\ra}{\rightarrow}

\newcommand{\bs}{\boldsymbol}

\newcommand{\SRO}{Sr$_2$RuO$_4$}
\newcommand{\bk}{\boldsymbol k}

\newcommand{\bkp}{\boldsymbol k^\prime}

\usepackage{times}

\begin{document}
\title{Higher-order topological superconductivity: possible realization in Fermi gases and \SRO}
\author{Zhigang Wu}
\email{wuzg@sustc.edu.cn}
\affiliation{Shenzhen Institute for Quantum Science and Engineering and Department of Physics,
Southern University of Science and Technology, Shenzhen 518055, China}
\author{Zhongbo Yan}
\email{yanzhb5@mail.sysu.edu.cn}
\affiliation{School of Physics, Sun Yat-sen University, Guangzhou 510275, China}
\author{Wen Huang}
\email{huangw001@gmail.com}
\affiliation{Center for Quantum Computing, Peng Cheng Laboratory, Shenzhen 518005, China}
\affiliation{Institute for Advanced Study, Tsinghua University, Beijing 100084, China}
\date{\today}

\begin{abstract}

We propose to realize second-order topological superconductivity in bilayer spin-polarized Fermi gas superfluids. We focus on systems with intralayer chiral $p$-wave pairing and with tunable interlayer hopping and interlayer interactions. Under appropriate circumstances, an interlayer even-parity $s$- or $d$-wave pairing may coexist with the intralayer $p$-wave. Our model supports localized Majorana zero modes not only at the corners of the system geometry, but also at the terminations of certain one-dimensional defects, such as lattice line defects and superfluid domain walls. We show how such topological phases and the Majorana zero modes therein can be manipulated in a multitude of ways by tuning the interlayer pairing and hopping. Generalized to spinful systems, we further propose that the putative $p$-wave superconductor Sr$_{2}$RuO$_{4}$, when subject to uniaxial strains, may also realize the desired topological phase.

\end{abstract}

\maketitle

Topological superconductors (TSCs) have been actively pursued in the last decade as they harbor gapless Majorana modes protected by bulk-boundary correspondence, on their one-dimensional ($1$D) lower boundaries or certain bulk defects \cite{qi2011,alicea2012new,Beenakker2013,stanescu2013majorana, leijnse2012introduction,Elliott2015,sarma2015majorana,sato2016majorana,aguado2017,Chiu2016RMP}. Protected by a nontrivial topological invariant, the zero-dimensional ($0$D) Majorana modes localized at the boundaries of $1$D TSCs \cite{Kitaev2001unpaired,lutchyn2010,oreg2010helical}, vortices or lattice dislocations of $2$D TSCs \cite{Read2000,fu2007c,Teo2010}, namely, the Majorana zero modes (MZMs), can realize nonlocal qubits immune to local decoherence and are thus expected to form the building blocks of topological quantum computation\cite{ivanov2001,Kitaev2003,Nayak2008,Alicea2011nonabelian}. Meanwhile, $1$D gapless Majorana modes or even higher dimensional gapless Majorana modes are dissipationless in transport along the free-moving directions. Consequently, they have interesting quantized responses to external probes \cite{Read2000,Wang2011,Nomura2012} and allow applications in, e.g., transport of heat. Remarkably, quantized signatures of $0$D MZMs and $1$D chiral Majorana modes have recently been reported in experiments \cite{Zhang2018quantized,He2017chiral,Banerjee2018,Kasahara2018}, signaling that this field is about to enter a new era.

Very recently, an important theoretical progress in this field is the recognition that the $m$D gapless Majorana modes can also emerge in an $n$D superconductor with $m\leq n-2$, even though the $(n-1)$D boundary of the superconductor is fully gapped \cite{TeoJ2013,Benalcazar2014,Yan2018hopf,Chan2017,Langbehn2017hosc,Khalaf2018hosc,Geier2018hosc,Zhu2018hosc,Yan2018hosc,Wang2018hosc2,Liu2018hosc,Wang2018hosc,Hsu2018,You2018hosc}.
We note that similar physics has also been explored in the context of insulating systems \cite{benalcazar2017quantized,Benalcazar2017prb,SongZD2017,Schindler2018HOTI}. Superconductors with such a novel topological property have been dubbed ``(n-m)$^{th}$-order topological superconductors'', or ``higher-order topological superconductors'' (HOTSCs), and have attracted increasing attention as they greatly enrich the boundary physics of superconductors as well as the platforms for obtaining gapless Majorana modes. Thus far, this topological phase has only been proposed in a few systems, including helical $p$-wave superconductor under a magnetic field \cite{Langbehn2017hosc,Zhu2018hosc}, topological insulator/high-temperature superconductor heterostructure \cite{Yan2018hosc,Wang2018hosc2,Liu2018hosc}, and helical $p$-wave superconductor/d-wave superconductor heterostructure \cite{Wang2018hosc}. All of these proposals focus on electronic systems and most of them rely heavily on the proximity effect.

Topological phases have also been actively sought after in cold atomic systems~\cite{cooper2018,zhang2018}, which have unparalleled advantages in controllability and tunability. For instance, both of the celebrated Su-Schrieffer-Heeger model~\cite{Su1979} and the Haldane model~\cite{Haldane1988}, two textbook models of the topological band theory, have been realized in cold atomic systems and the associated topological phase transitions explored~\cite{Atala2013experimental,jotzu2014experimental}.
This motivates us to introduce the concept of higher-order topological superfluid (HOTSF), the neutral counterpart of HOTSC, in degenerate Fermi gases. In this paper, we show that a bilayer spin-polarized Fermi gas with intralayer chiral $p$-wave pairing and interlayer even-parity pairing provides a viable platform to realize intrinsic second-order topological superfluids (SOTSF). MZMs could emerge, not only at the corner of the system geometry, but also at the end points of certain 1D defects, such as lattice line defects and superfluid domain walls. A remarkable advantage of our proposal, as we shall demonstrate, lies in the ease of tuning the system across distinct topological phases.

Generalized to spinful systems, we further propose a pristine material platform for realizing the desired topological superconductivity -- \SRO~\cite{Maeno1994}. This material has long been hailed a candidate spin-triplet $p$-wave superconductor \cite{Rice1995,Baskaran1996,Maeno2001,Mackenzie2003,Kallin2009,Kallin2012,Kallin2016,Liu2015,Mackenzie17,Huang2018}. Intriguingly, recent experimental signatures indicative of spin-singlet pairing at large uniaxial strains \cite{Hicks2014,Steppke2017,Watson2018} raises the prospect to continuously drive its superconducting state from one pairing symmetry to another. It is therefore sensible to conjecture a region of mixed-parity pairing at intermediate strains. We will discuss possible experimental means to identify this phase.

{\it Bilayer Fermi gas superfluid.---} Let us first introduce the bilayer spin-polarized Fermi gas system and investigate its phase diagram.
For concreteness, we assume the fermions in each layer move in a square optical lattice potential and are described by a tight-binding model with dispersion $\xi_{\bk} = -2 t \left (\cos k_x + \cos k_y \right) -\mu$, where $\bk = (k_x, k_y)$, $t$ is the in-plane hopping, and $\mu$ sets the chemical potential. In addition, fermions on the two layers can hybridize with amplitude $t_\perp$. Now, each layer of the spin-polarized Fermi gas can form a topological $p_x\pm ip_y$ superfluid, either through a stable $p$-wave Feshbach resonance~\cite{Regal2003,Ticknor2004,Esslinger2005,Gurarie2005,Zhang2004,Schunck2005,Chevy2005,Inada2008,Nakasuji2013} or through an induced attractive interaction using atomic mixtures~\cite{Viverit2000,Wu2016,Kinnunen2018}. For two layers of such $p$-wave superfluid, additional interlayer even-parity pairing channels, such as $s$- or $d$-wave, can be established through relevant Feshbach resonances. The system is described by the Bogoliubov de-Gennes (BdG) Hamiltonian
\begin{eqnarray}
\hat H = \sum_{\bk, i,j} \left\{ \epsilon_{ij}(\bk)  c^\dag_{\bk i}  c_{\bk j} + \frac 1 2\left[ \Xi_{ij}(\bk) c_{\bk i}  c_{-\bk j} +h.c.\right]  \right\},\label{BdG}
\end{eqnarray}
where $i,j = 1,2$ is the layer index, $c_{\bk i}$ creates a fermion with quasi-momentum $\bk$ in layer $i$, $\epsilon_{jj}(\bk)  = \xi_{\bk}$ and $\epsilon_{ij}(\bk)= t_{\perp}$ for $i\neq j$. We separate out the form factors describing the symmetry of the pairings, viz. $\Xi_{jj}(\bk) = \Delta_{jj}f_{j\bk}$, and $\Xi_{21}(\bk)=\Xi_{12}(\bk)= \Delta_{12} g_{\bk}$, where $f_{j\bk} = \sin k_x \pm i\sin k_y $ describes the $p$-wave pairing and $g_{\bk}$ is an even-parity function corresponding to the $s$-wave or $d$-wave pairing. The pairing gaps are determined by the following coupled gap equations,
\begin{align}
\Xi_{ij}(\bk) = -\sum_{\bkp} V_{ij}(\bk,\bkp)\langle  c_{\bkp i}   c_{-\bkp j} \rangle,
\label{gapeq}
\end{align} where $V_{ii}(\bk,\bkp) = -V_{p} f_{i\bk}f^\ast_{i\bkp}$ and $V_{12}(\bk,\bkp) = -V_{s(d)} g_{\bk}g_{\bkp}$ denote the intra- and interlayer pairing interactions, respectively.

As no symmetry requires or precludes the said intra- and interlayer pairings to coexist, such a state may only emerge in a limited interaction parameter space. A previous study on a similar model~\cite{Midtgaard2017}, considering interlayer $s$-wave pairing and with vanishing $t_\perp$, found indeed regions of coexistence where the $p$-wave pairings on the two layers preferentially develop opposite chirality (analogous to the $^3$He B-phase \cite{Vollhardt1990}), rather than same chirality. Hereafter, we refer to them respectively as helical $p$-wave and chiral $p$-wave pairings, in reference to the notion in spinful models \cite{Vollhardt1990}. Importantly, here we show that, such a state is stable against sizable interlayer hopping, and the coexistence with other types of interlayer even-parity pairings, such as $d$-wave, is also possible.

\begin{figure}
\subfigure{\includegraphics[width=4.25cm]{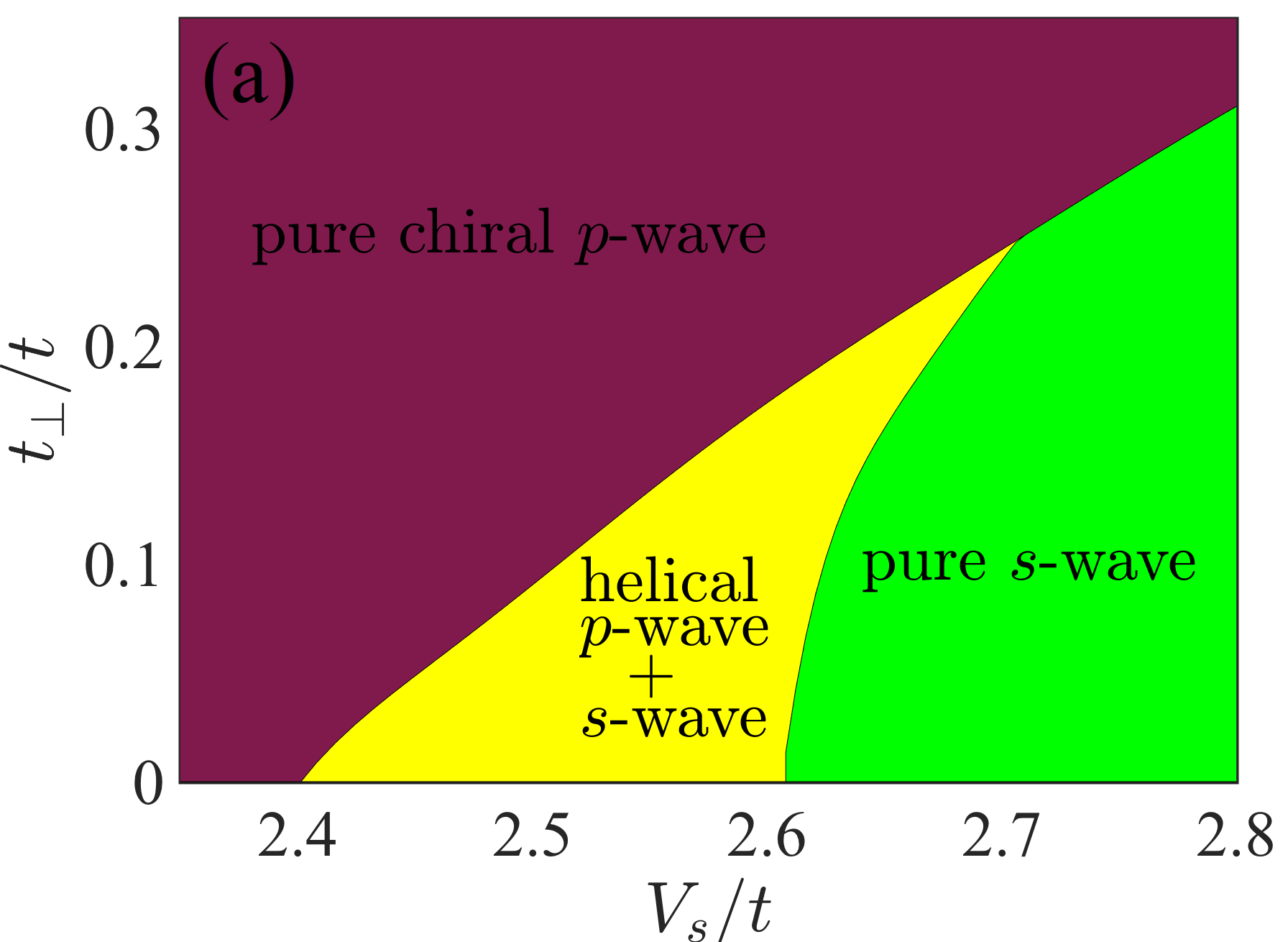}}
\subfigure{\includegraphics[width=4.25cm]{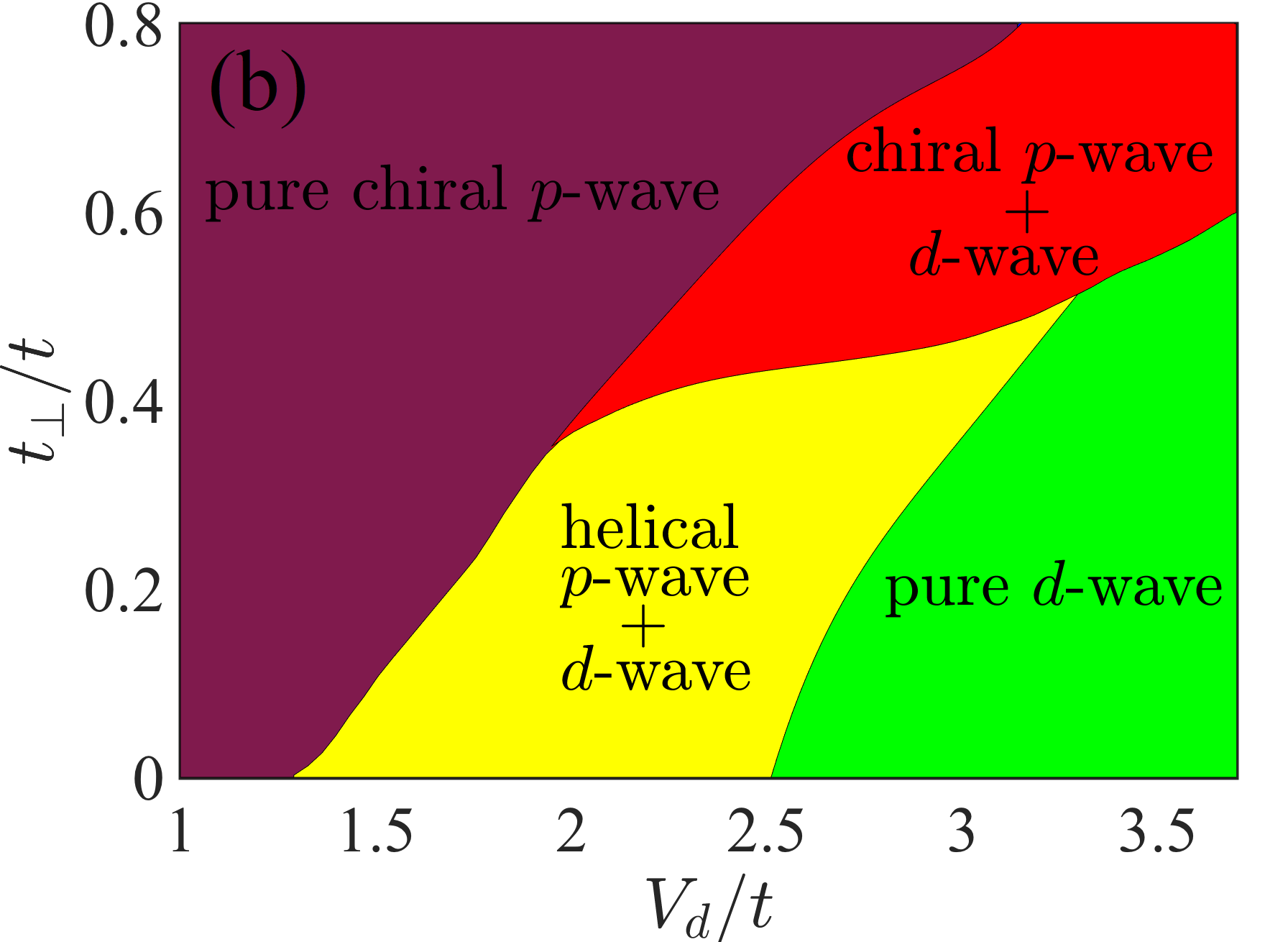}}
\caption{(color online) Mean-field phase diagrams of the bilayer model with intralayer chiral $p$-wave pairing $f_{j\bk}=\sin k_x \pm i\sin k_y$ and interlayer pairing in the: (a) $s$-wave $g_{\bk}=1$ and (b) $d$-wave $g_{\bk}=\cos k_x - \cos k_y$ channels. Calculations were performed with $\mu=-0.5t$. The $p$-wave interaction strength $V_p$ is taken to be $2.5t$ in (a) and $2.2t$ in (b). The states with the same and opposite chiral $p$-wave pairings on the two layers are referred to as helical $p$-wave and chiral $p$-wave pairings, respectively. }
\label{fig:phaseDiag}
\end{figure}

Figure \ref{fig:phaseDiag} shows the representative phase diagrams obtained from our self-consistent BdG solutions of Eq.~(\ref{gapeq}). Qualitatively similar behavior is seen for both interlayer $s$-wave and $d$-wave pairings. In particular,  at smaller interlayer hopping and where intra- and interlayer pairings coexist (i.e. a mixed-parity phase), the helical $p$-wave pairing is always more stable. This can be understood by inspecting the following terms in the free energy \cite{supplemental},
\begin{equation}
\beta^\prime \left( \Delta^\ast_{11}\Delta^\ast_{22}\Delta_{12}^2 + c.c.  \right)
\label{eq:f4a}
\end{equation}
Here, $\beta^\prime$ is  proportional to the Fermi surface average of the quantity $-f^\ast_{1\bk} f^\ast_{2\bk} g_{\bk}^2$. Hence, $\beta^\prime<0$ for helical pairing and $\beta^\prime=0$ for chiral pairing. This term is thus minimized in the helical state, along with the phases of $(\Delta_{11},\Delta_{22},\Delta_{12})$ taking, e.g. $(1,1,\pm 1)$. The choice of ``$\pm 1$" implies ground state degeneracy, whose concomitant $Z_2$ symmetry allows for the formation of superfluid domains. The system in fact possesses an exact $U(1)\times U(1)$ symmetry, manifested in an added freedom to choose the relative phase between the order parameters \cite{supplemental}. This originates from the conservation of Cooper pair angular momentum in the free energy, which dictates that any phase-dependent coupling must appear with the elementary unit of $\left( \Delta^\ast_{11}\Delta^\ast_{22}\Delta_{12}^2 + c.c.  \right)$, up to all higher orders. Nonetheless, we shall take the assumption that one particular phase configuration is stabilized, due to a coupling to some unspecified external sources. It is worth stressing that these conclusions hold irrespective of the detailed form of the interlayer even-parity pairing \cite{footnote1}. We also note a similar emergent $U(1)\times U(1)$ symmetry in a somewhat different context \cite{WangYX2017}.

As shown in Fig. \ref{fig:phaseDiag}, increasing interlayer hopping reduces the region of the coexistence and tips the balance in favor of the pure chiral $p$-wave pairing. This is mainly because the chiral state benefits from an induced Josephson coupling $\propto -t_\perp^2 \Delta_{11}^\ast \Delta_{22} + c.c.$, whilst the helical state does not. In addition, the interlayer mixing acts as an effective Zeeman field which can further suppress the interlayer pairing.

Lastly, we briefly discuss the experimental realization of such a bilayer Fermi gas system. Although all the ingredients of our proposal are within current experimental reach, a fermionic $p$-wave superfluid is yet to be realized. The central challenge in achieving such a superfluid is the engineering of a strong $p$-wave interaction so that the superfluid transition temperature is not prohibitively small. Two routes can be pursued for this purpose. One is through the $p$-wave Feshbach resonances, which have been experimentally explored for both $^{40}$K~\cite{Regal2003,Ticknor2004,Esslinger2005} and $^6$Li~\cite{Zhang2004,Schunck2005,Chevy2005,Inada2008,Nakasuji2013} atoms. Unfortunately the life times for these resonances are found to be relatively short due to the severe three-body losses. However, a recent study shows that by modulating the depth of the optical lattice the inelastic collisional losses can be significantly suppressed~\cite{Fedorov2017}. This raises the prospect that a strong and stable $p$-wave interaction can be achieved in moving closer to the Feshbach resonances. Another method is to utilize induced interactions in atomic mixtures~\cite{Wu2016,Kinnunen2018}. In such a scheme, a layer of spin-polarised non-interacting Fermi gas is immersed in the Bose-Einstein condensate and gains an effective attraction through exchanging the phonons in the Bose gas. By increasing the Bose-Fermi coupling, it is possible to generate a strong effective p-wave interaction between the fermions without destabilizing the system. In fact, very recently a mixed-dimensional $^{174}$Yb-$^7$Li mixture has been created in experiment~\cite{Takahashi2018}, paving the way for the realization of a 2D $p$-wave superfluid.

{\it Second-order topological superfluids and corner MZMs.---}
Let us now focus on the coexistence region with the intralayer helical $p$-wave pairing and the interlayer even-parity pairing. To gain an intuitive understanding of the topological property of this region, we consider the continuum Hamiltonian by expanding the lattice Hamiltonian in Eq.(\ref{BdG}) around $\bk=(0,0)$. Introducing the spinor $\Psi_{\bs k} = (c_{\bs k 1}, c_{\bs k 2}, c^{\dagger}_{-\bs k1}, c^{\dagger}_{-\bs k 2})^T$, the expansion returns $\hat H =\frac{1}{2} \int d\bk\Psi_{\bs k}^\dag H(\bk) \Psi_{\bs k}$ with
\begin{eqnarray}
H(\bk)&=&[t(k_{x}^{2}+k_{y}^{2})-m_{0}]\tau_{z}+t_{\perp}\tau_{z}s_{x}\nonumber\\
&&+\Delta_{p}(k_{x}\tau_{x}-k_{y}\tau_{y}s_{z})-\Delta_{e}(\bk)\tau_{y}s_{y},
\end{eqnarray}
where $\tau_{i}$ and $s_{i}$ ($i=x,y,z$) are Pauli matrices acting on the particle-hole
space and the bilayer space, respectively; $m_{0}=4t+\mu$ and $\Delta_{e}(\bk)=\Delta_{s}-\Delta_{d}(k_{x}^{2}-k_{y}^{2})/2$. Here we take $\Delta_{11}=\Delta_{22}=\Delta_p$. For general purposes, we have incorporated both $s$-wave and $d$-wave pairings in the interlayer pairing $\Delta_{e}(\bk) \equiv \Xi_{12}(\bk)$ where ``e'' stands for even parity.

Without loss of generality, in the following we consider positive $t$ and $m_{0}$, so that for vanishing $t_{\perp}$ and $\Delta_{e}$, the Hamiltonian describes a symmetry-protected topological superfluid with helical edge states~\cite{Midtgaard2017}. The protecting symmetry is a pseudo time-reversal symmetry associated with the operator $\mathcal{T}=i\tau_{z}s_{y}\mathcal{K}$ ($\mathcal{K}$ denotes complex conjugate). The presence of interlayer hopping and pairing in general breaks this symmetry, introducing Dirac mass terms to gap out the helical edge states. To understand how a SOTSF is realized, below we turn to an effective edge theory.

\begin{figure}
\includegraphics[width=8cm]{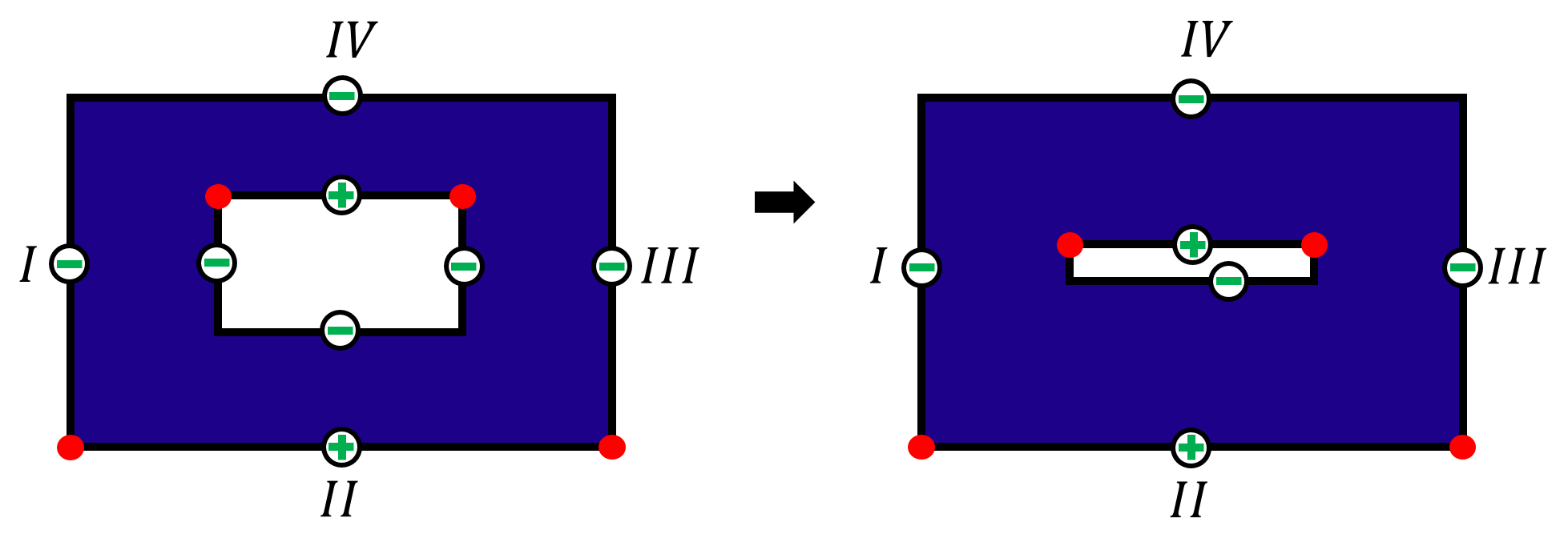}
\caption{(color online) Top view of a finite-size bilayer system with a depleted patch (left) and a line defect (right). The four outer edges are labeled I, II, III, and IV. The line defect can be viewed as a depleted patch squeezed to one dimension. The symbols in circles represent the sign of the effective Dirac masses on the edges of a particular mixed-parity state. The red dots denote the kinks where the Dirac mass switches sign, i.e. where the MZMs are bounded. }
\label{fig:defectDiag}
\end{figure}

We label the four outer edges of a square lattice I, II, III and IV (see Fig.\ref{fig:defectDiag}) and define a 1D ``boundary coordinate'' $l$ stretching these edges in a counterclockwise fashion. To further simplify the analysis, we treat $t_{\perp}$ and $\Delta_{e}$ as small perturbations. Following the analyses in Ref.\onlinecite{Yan2018hosc} and as explained in more detail in the supplemental material~\cite{supplemental}, we obtain an effective 1D Dirac Hamiltonian,
\begin{eqnarray}
H_{\rm edge}(l)=iv_{l}\partial_{l}s_{z}+M_{l}s_{y},
\end{eqnarray}
where the velocity $v_{l}$ and the Dirac mass $M_{l}$ are defined on the four segments of the 1D coordinate as follows: $v_{l}\equiv\Delta_{p}$, and
$M_{l}=\tilde{\Delta}_{d}-\Delta_{s}, -t_{\perp}-\tilde{\Delta}_{d}-\Delta_{s}, \tilde{\Delta}_{d}-\Delta_{s}, t_{\perp}-\tilde{\Delta}_{d}-\Delta_{s}$, for $l=$ I, II, III, IV, respectively. Here $\tilde{\Delta}_{d}=\Delta_{d}m_{0}/t$. An interesting observation is that $t_{\perp}$ enters only in $M_{\rm II}$ and $M_{\rm IV}$, which suggests a selective influence of the interlayer hopping on different edges. This can be simply understood from the fact that while the interlayer hopping term anti-commutes with $k_x\tau_x$, it commutes with $k_y\tau_y s_z$.

Let us first focus on the case where interlayer pairing is purely $d$-wave. The Dirac fermion acquires a mass of
$M_{l}=\tilde{\Delta}_{d}, -t_{\perp}-\tilde{\Delta}_{d}, \tilde{\Delta}_{d}, t_{\perp}-\tilde{\Delta}_{d}$ on the four respective edges. Without the interlayer hopping, it is readily seen that $M_l$ changes sign at every corner and thus each corner forms a kink. According to the Jackiw-Rebbi theory \cite{Jackiw1976b}, each kink hosts one zero mode. As a consequence, the coexisting helical $p$-wave and $d$-wave pairings constitute a SOTSF with one MZM per corner, consistent with recent studies \cite{Wang2018hosc}. interlayer hopping introduces richer phases. In particular, when $|t_{\perp}|$ exceeds $|\tilde{\Delta}_{d}|$, the number of kinks reduces to two, indicating that the system undergoes a topological phase transition and turns into a different SOTSF with only two corner MZMs. We stress that the requirement for interlayer hopping to be stronger than the interlayer pairing is well within reach for a weak-coupling superfluid.

\begin{figure}
\includegraphics[width=8.cm]{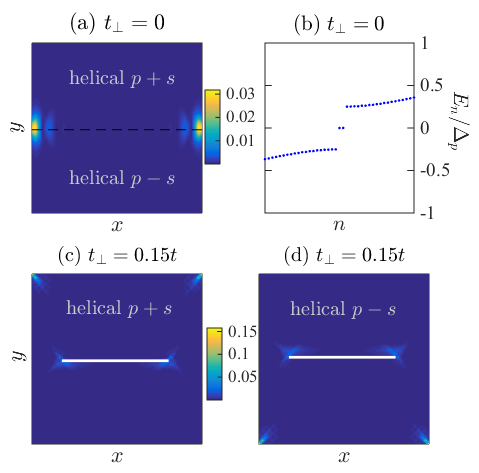}
\caption{(color online) Upper panel: (a) density distribution of a pair of MZMs localized at the terminations of superfluid domains of the bilayer model with mixed-parity pairing and with $t_\perp=0$, and (b) the corresponding low-energy spectrum. The dashed line in (a) marks the domain wall separating the indicated superfluid domains. (c) and (d),  density distribution of a pair of corner MZMs and another pair of MZMs at the ends of a line defect for $t_\perp=0.15t$. The narrow depleted patches in the middle of (c) and (d) designate the line defects. They are simulated by removing the relevant lattice sites, along with the associated pairings and hoppings across the defect, from the Hamiltonian. These BdG calculations were performed on the same lattice model as in Fig. \ref{fig:phaseDiag}, in an open-boundary geometry with a system size of $50\times 50$ . The amplitudes of the intralayer helical $p$-wave and interlayer $s$-wave pairings are taken to be $|\Delta_{11}|=|\Delta_{22}|=\Delta_p=0.4t$, $|\Delta_{12}|=\Delta_{s}=0.1t$, respectively. }
\label{fig:cornerstate}
\end{figure}

We now proceed to the  $s$-wave case. The Dirac masses at the four edges now take the following values
$M_{l}=-\Delta_{s}, -t_{\perp}-\Delta_{s}, -\Delta_{s}, t_{\perp}-\Delta_{s}$. Without the interlayer hopping $M_l$ is uniform across the boundary, thus the kinks and the concomitant corner MZMs are absent. However, this phase is not a featureless trivial superfluid. As mentioned above, the mixed-parity state has a nontrivial $Z_{2}$ symmetry associated with two degenerate ground states, which allows for the formation of superfluid domains. At the edge where two domains meet, such as that depicted in Fig. \ref{fig:cornerstate} (a), a new Dirac kink develops with masses of $\Delta_s$ and $-\Delta_s$ on the opposite sides. Therefore one MZM appears on this end of the domain wall, along with a partner mode on the other end, as demonstrated numerically for a lattice model in Fig. \ref{fig:cornerstate} (a) and (b). We note MZMs of the same origin in the interlayer $d$-wave model.

In the more interesting case where the system is free of domains, sign-changing Dirac masses are possible by making $|t_{\perp}|>|\Delta_{s}|$. Similar to the $d$-wave model above, $M_l$ changes sign at two of the corners and a pair of MZMs are formed [Fig. \ref{fig:cornerstate} (c)]. Remarkably, a $\pi$ phase change for $\Delta_s$ alternates the location of the kinks, and therefore that of the corner MZMs. As one can tell from Fig. \ref{fig:cornerstate} (c) and (d), MZMs appear at the two upper corners (i.e. two ends of edge IV) when $t_{\perp}\cdot\Delta_{s}>0$, and at the two lower corners (i.e. two ends of edge II) when $t_{\perp}\cdot\Delta_{s}<0$. Similar observation can be made for the $d$-wave model when $|t_\perp|>|\tilde\Delta_d|$.

For the case with non-vanishing $\Delta_{s}$ and $\tilde{\Delta}_{d}$ at the same time, the topological property can be analyzed similarly. Qualitatively speaking, when $\tilde{\Delta}_{d}$ dominates over $\Delta_{s}$ and $t_{\perp}$, the system supports four corner MZMs; when $t_{\perp}$ dominates, it carries a pair of MZMs at two of the corners; and when $\Delta_{s}$ dominates, MZMs are absent at lattice corners, but can still emerge if the system develops superfluid domains. Since MZMs can appear in all of the above cases, we designate these mixed-parity phases SOTSF. Note that our edge theory is not suited beyond a critical value of $t_\perp$ when the bulk becomes gapless and Majorana flatbands develop at certain edges. We shall not elaborate this scenario here. As an important final remark, the boundary Dirac kinks necessary for the formation of isolated MZMs persist for more general system geometries.

{\it MZMs at bulk line defects.---} The analyses above have demonstrated the extraordinary flexibility to manipulate the topological superfluid phase. We now turn to an interesting advantage of the SOTSF phase possessing only one pair of corner MZMs\cite{Langbehn2017hosc,Khalaf2018hosc,Geier2018hosc,Zhu2018hosc}, otherwise absent in the phase with MZMs at each of the four lattice corners \cite{Yan2018hosc,Wang2018hosc2,Liu2018hosc,Wang2018hosc}, namely, the emergence of MZMs at the end points of line defects in the bulk.

A line defect is created by removing from the Hamiltonian an array of lattice sites, as well as the associated pairing and hopping terms. In cold atom experiments, this may be achieved by shining an extra laser beam to create a strong local potential barrier. Our numerical simulation in Fig.\ref{fig:cornerstate}(c) and (d) indeed show a pair of MZMs at the ends of the line defects, resembling the scenario in an open Kitaev chain \cite{Kitaev2001unpaired}. The origin of these end MZMs can be understood as follows. Let us first imagine caving out a large rectangular patch in the bulk of the lattice [see Fig.\ref{fig:defectDiag}(a)], then the two corners of the inner boundary shall host MZMs according to our analyses above. We next reduce the width of the patch along the direction perpendicular to the edge hosting the two MZMs to make for a line defect [Fig.\ref{fig:defectDiag}(b)]. Since these MZMs do not couple with other high-energy states due to particle-hole symmetry, the enhanced confinement has no effect on them. This also manifests in the resilient Dirac mass configuration on the four inner edges during the process, as depicted in Fig. \ref{fig:defectDiag}. As a result, the two corner MZMs eventually become two end MZMs on the line defect. The same argument applies to more general line defects, except those oriented perpendicular to the line connecting the two corner MZM in Fig. \ref{fig:cornerstate} (c) and (d).

By contrast, in HOTSCs and HOTSFs with MZMs at every corner, compressing the depleted patch to a line inevitably brings together even number of corner MZMs and split their energies. Thus line defects in those phases do not support robust MZMs. Additionally, lattice dislocations, formed by ``gluing'' back the two sides of the line defect using the corresponding pairings and hoppings, do not support MZMs. This can be attributed to the destruction of the Dirac kinks by the said procedure. We observe that this contrasts with a separate scenario in Ref. \cite{Hugher2014}, where a nontrivial 1D $Z_2$ invariant \cite{Kitaev2001unpaired} protects the MZMs bound to the dislocations.

{\it Mixed-parity pairing in} \SRO.--- At $t_\perp=0$, the above bilayer model is an exact dual to a single-layer spinful model with a mixture of spin-triplet helical $p$-wave and spin-singlet even-parity pairings. In the most general form, the gap function reads $\hat{\Xi}_{\bs k} = (\Delta_{p} \bs d_{\bs k} \cdot \bs \sigma + \Delta_e g_{\bs k})i\sigma_y$, where $\sigma_i$'s are Pauli matrices, $\bs d_{\bs k}$ represents the basis function of the helical $p$-wave (such as $k_x \hat{y} \pm k_y \hat{x}$ according to standard notations \cite{Vollhardt1990}). A possible realization by proximitizing $p$-wave and $d$-wave superconductors has indeed been recognized and analyzed in detail in a recent work \cite{Wang2018hosc}. We note that in accordance with the situation in the bilayer model, $\Delta_p$ and $\Delta_e$ must differ by a phase of $\pm \pi/2$ due to the peculiar form of $\bs d_{\bs k}$.

Here, we propose that a pristine material platform, \SRO, may realize this topological phase without involving proximity effects. Widely considered a $p$-wave superconductor, this compound yet displays characters suggestive of spin-singlet pairing in the presence of large uniaxial strains \cite{Hicks2014,Steppke2017}. Further, there are also implications of weak out-of-plane magnetic field favoring helical $p$-wave pairing \cite{Murakawa2004,Annett2008,Ueno2013} in unperturbed \SRO. Incidentally, microscopic theoretical analyses often find, in certain regimes of parameter space, leading helical or even-parity superconducting channels \cite{Scaffidi2014, Huang2016, Zhang2017b, WangQH2016, Kim2017, LiuYC2018,Steffens2018,Gingras2018}.  It is therefore plausible that a mixed-parity phase with coexisting helical and even-parity pairings emerges in the presence of an intermediate strain and a weak out-of-plane field, as we elaborate in Ref. \onlinecite{supplemental}. The local density of states (LDOS) at certain representative locations of the sample geometry can serve as diagnosis of this phase. Take the example where corner MZMs are formed, the corner LDOS shall exhibit a sharp zero-bias peak, while other sample locations must instead see gapped spectra. An illustration is provided in Ref. \onlinecite{supplemental}.

{\it Summary.---} We have shown that a bilayer spin-polarized Fermi gas with intralayer chiral $p$-wave and interlayer even-parity pairings provides a feasible platform to realize a variety of SOTSFs, each supporting MZMs at certain corners and/or 1D line defects of the system geometry. In particular, by manipulating the interlayer pairing and interlayer hopping, it is possible to drive the transition across multiple distinct SOTSFs. Further, among the possible topological phases, the one with two corner MZMs possesses an unparalleled advantage that their line defects in the bulk host robust MZMs. This allows for potential applications in, as a typical example, the design of MZM circuits for quantum computation in a single system \cite{Karzig2017}. Generalized to spinful systems, we also proposed that the uniaxially strained \SRO~may be driven into a mixed-parity phase, thereby realizing the desired higher-order topological superconductivity. Given the recent progress on the uniaxial strain experiments \cite{Hicks2014,Steppke2017}, this proposal represents a particularly promising route.

{\it Acknowledgements.---} We would like to acknowledge helpful discussions with Manfred Sigrist and Hong Yao. Z.W. acknowledges the support by the Science, Technology and Innovation Commission of Shenzhen Municipality and Guangdong Innovative and Entrepreneurial Research Team Program. Z.Y. acknowledges the support by a startup grant at Sun Yat-sen University. W.H. is supported by the C. N. Yang Junior Fellowship of the Institute for Advanced Study at Tsinghua University.

\bibliography{dirac}

\begin{thebibliography}{98}
\bibitem [{foo()}]{footnote1}%
  \BibitemOpen
  \href@noop {} {}\bibinfo {note} {Interlayer odd-parity pairing similarly
  leads to opposite-chirality state. However, the resultant state does not
  support corner Majorana modes.}\BibitemShut {Stop}%
\end{thebibliography}

\widetext
\clearpage
\begin{center}
\textbf{\large Supplemental Material for ``Higher-order topological superconductivity: possible realization in Fermi gases and \SRO"}\\
\vspace{4mm}
{Zhigang Wu$^{1,*}$, Zhongbo Yan$^{2,\dag}$ and Wen Huang$^{3,\ddag}$}\\
\vspace{2mm}
{\em \small $^1$Shenzhen Institute for Quantum Science and Engineering and Department of Physics, \\
Southern University of Science and Technology, Shenzhen 518055, China \\
$^2$School of Physics, Sun Yat-sen University, Guangzhou, 510275, China \\
$^3$Institute for Advanced Study, Tsinghua University, Beijing, 100084, China}
\end{center}

\setcounter{equation}{0}
\setcounter{figure}{0}
\setcounter{table}{0}
\makeatletter
\renewcommand{\theequation}{S\arabic{equation}}
\renewcommand{\thefigure}{S\arabic{figure}}
\renewcommand{\bibnumfmt}[1]{[S#1]}

This supplemental material contains the following four sections:  (I) Solutions of the gap equation and the phase diagram; (II) Ginzburg-Landau theory; (III) Effective edge theory; (IV) Mixed-parity pairing in \SRO.

\section{I. Solutions of the gap equation and the phase diagram}
Introducing $\Psi_{\bs k} = (c_{\bs k 1}, c_{\bs k 2}, c^{\dagger}_{-\bs k1}, c^{\dagger}_{-\bs k 2})^T$, the BdG Hamiltonian in Eq.~(1) of the main text  can be written as
\begin{equation}
\hat H = \frac{1}{2} \sum_{\bs k}
\Psi^\dagger_{\bs k}
\mathcal{H}_{\bs k}
\Psi_{\bs k},
\end{equation}
where
\begin{equation}
\mathcal{H}_{\bs k} =
\begin{bmatrix}
\xi_{\bs k} & t_\perp & \Xi_{11}(\bs k) &   \Xi_{12}(\bs k) \\
t_\perp & \xi_{\bs k} & -\Xi_{12}(\bs k) &   \Xi_{22}(\bs k)  \\
 \Xi^*_{11}(\bs k)  & -\Xi^*_{12}(\bs k) & -\xi_{\bs k}  & -t_\perp \\
\Xi^*_{12}(\bs k) &\Xi^*_{22}(\bs k)   & -t_\perp & -\xi_{\bs k}
\end{bmatrix}.
\label{eq:BdGham}
\end{equation}

\begin{figure}[b]
\subfigure{\includegraphics[width=8cm]{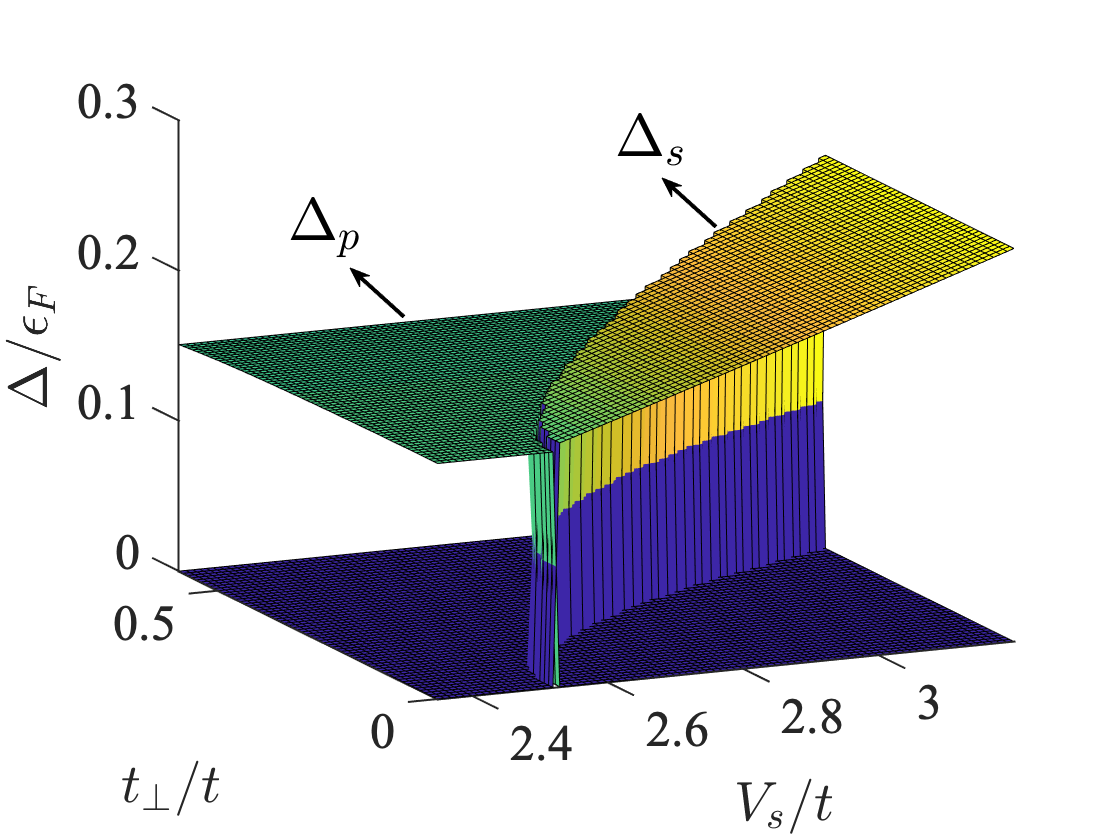}}
\subfigure{\includegraphics[width=8cm]{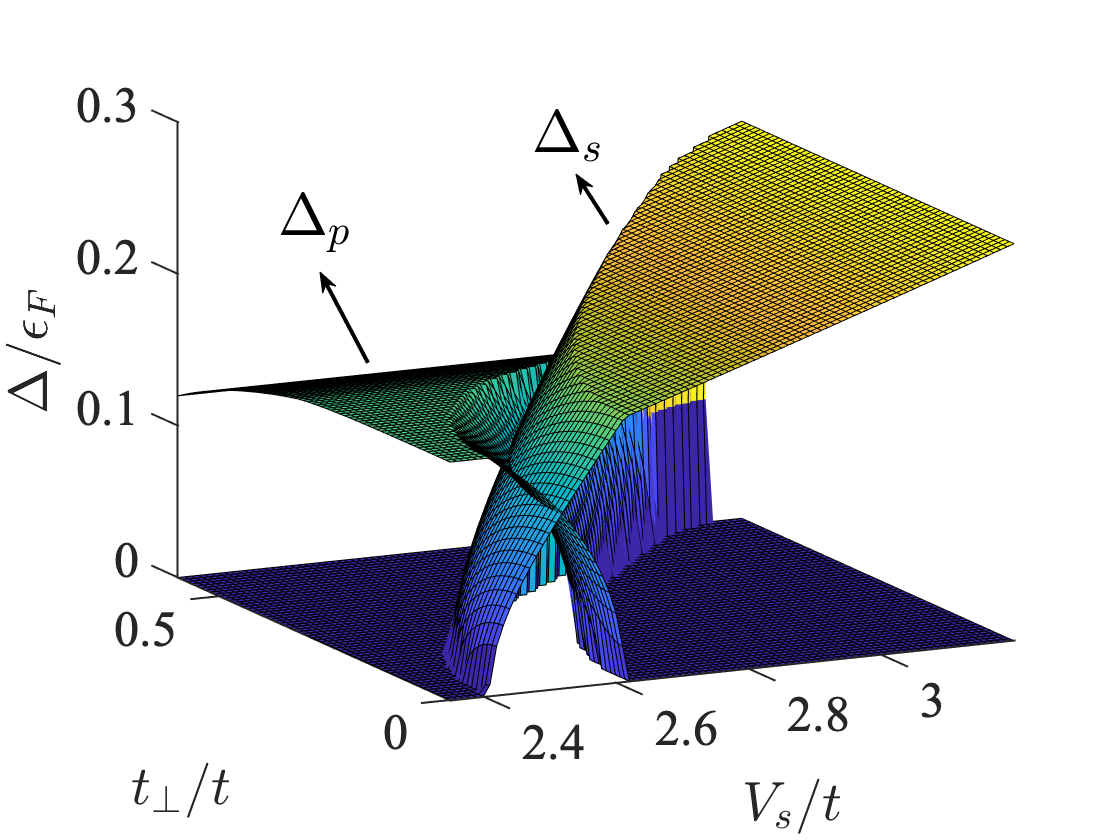}}
\caption{(color online) Solutions of the gap equations in the presence of the $s$-wave interlayer pairing interaction. Here $V_p = 2.5 t$, $\Delta_p \equiv |\Delta_{11}|$ and $\Delta_{s}\equiv |\Delta_{12}|$, measured against the Fermi energy $\epsilon_F =\mu + 4t $. Left:  The intralayer $p$-wave pairings have the same chirality. Right: The intralayer $p$-wave pairings have the opposite chirality. }
\label{fig:swavegap}
\end{figure}
\begin{figure}[h]
\subfigure{\includegraphics[width=8cm]{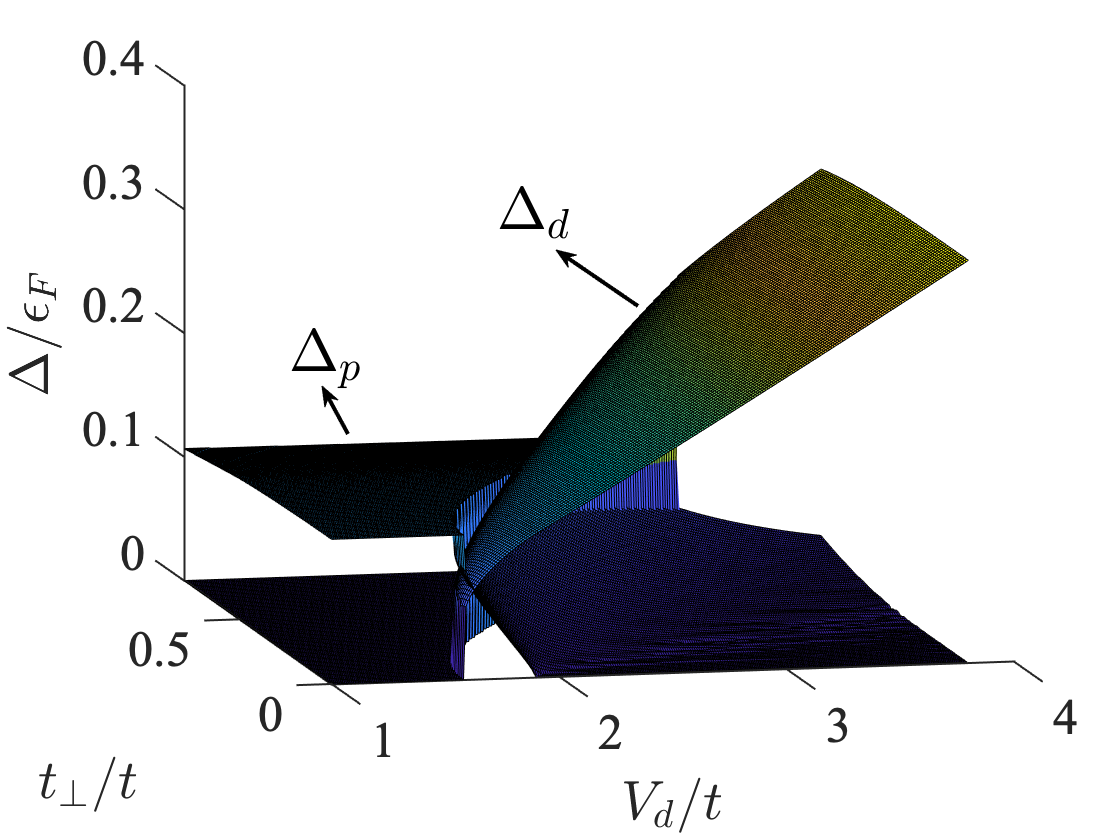}}
\subfigure{\includegraphics[width=8cm]{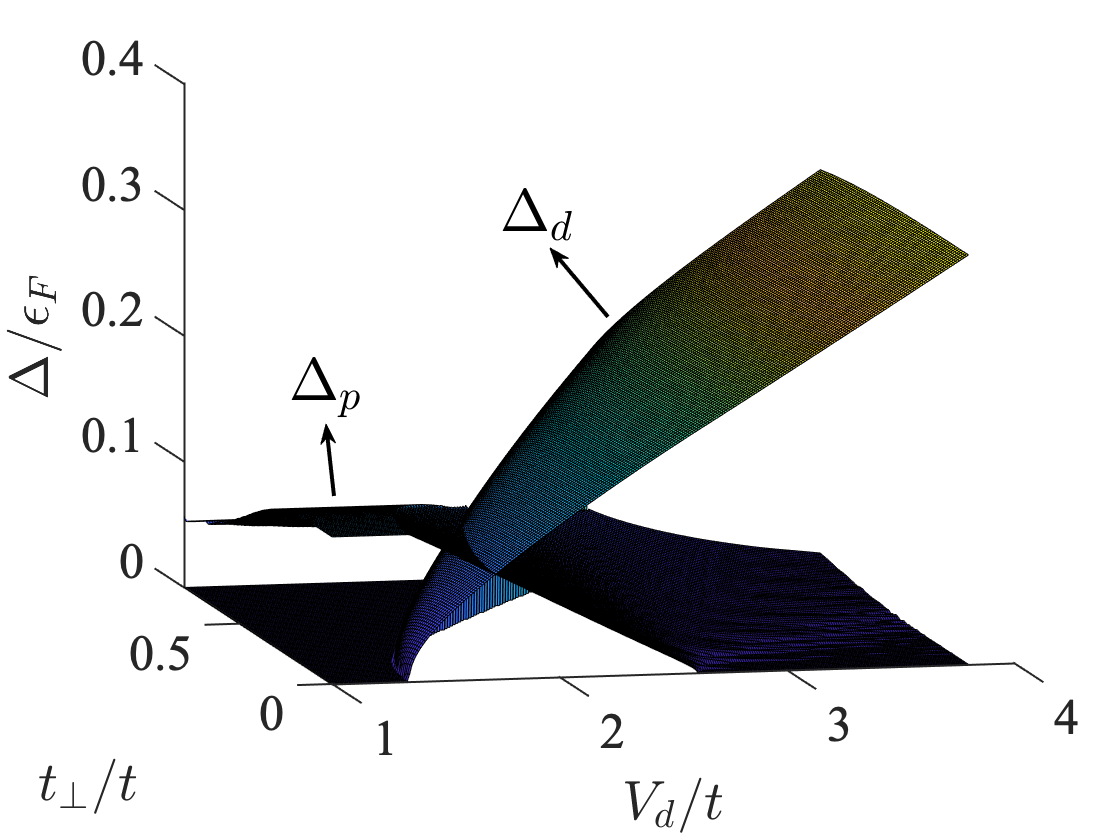}}
\caption{(color online)  Solutions of the gap equations in the presence of the $d$-wave interlayer pairing interaction. Here $V_p = 2.2 t$, $\Delta_p \equiv |\Delta_{11}|$ and $\Delta_{d}\equiv |\Delta_{12}|$, measured against the Fermi energy $\epsilon_F =\mu + 4t $. Left:  The intralayer $p$-wave pairings have the same chirality. Right: The intralayer $p$-wave pairings have the opposite chirality.}
\label{fig:dwavegap}
\end{figure}
The above Hamiltonian can be diagonalized by the following Bogoliubov transformation
\begin{align}
c_{\bs k j}  = u_{j+}(\bs k) b_{\bs k,+} + v_{j+}^* (\bs k) b^\dag_{-\bs k,+}  + u_{j-}(\bs k) b_{\bs k,-} + v_{j-}^* (\bs k) b^\dag_{-\bs k,-},
\end{align}
where $u_{j\pm},v_{j\pm}$ are the Bogoliubov amplitudes and $b_{\bs k,\pm}$ are the quasi-particle operators. The Bogoliubov amplitudes are obtained from the eigenvalue equation
\begin{align}
\mathcal{H}_{\bs k} \chi_{\bs k,\pm} = E_{\bs k, \pm} \chi_{\bs k,\pm},
\label{Heig}
\end{align} where $\chi_{\bs k,\pm} \equiv [u_{1\pm}(\bs k) , u_{2\pm}(\bs k), v_{1\pm}(-\bs k), v_{2\pm}(-\bs k) ]^T$. Using the transformation and the relevant ansatz for the pairing functions, i.e., $\Xi_{jj}(\bk) = \Delta_{jj}f_{j\bk}$, and $\Xi_{12}(\bk)= \Delta_{12} g_{\bk}$, the gap equation (2) of the main text can be written as
\begin{align}
\Delta_{11} &= -V_p \sum_{\bk} f^*_{1\bs k}\left [ u_{1+}(\bs k)v_{1+}^*(\bs k) + u_{1-}(\bs k)v_{1-}^*(\bs k)\right ]  \\
\Delta_{12} &= -V_{s(d)} \sum_{\bk} g^*_{\bs k}\left [ u_{1+}(\bs k)v_{2+}^*(\bs k) + u_{1-}(\bs k)v_{2-}^*(\bs k)\right ].
\label{gapeq1}
\end{align}

The amplitude of the gap functions $\Delta_{ij}$ are determined by solving Eqs.~(\ref{Heig}) and (\ref{gapeq1}) self-consistently for a specific chemical potential . For the intralayer $p$-wave pairing,  two types of solutions are possible, i.e., $\Xi_{22} = \Xi_{11}$ and $\Xi_{22} = \Xi^*_{11} $, corresponding to the same-chirality (chiral) and opposite-chirality (helical) solutions respectively.  Solutions for both of these cases are respectively shown in Fig.~\ref{fig:swavegap} and \ref{fig:dwavegap} for $\mu = -0.5t$.  We note that in the presence of the interlayer $s$-wave pairing interaction, the coexistence of the intra- and interlayer pairing occur only when the intralayer pairing is helical. This is in contrast to the $d$-wave case, where the coexistence is allowed for both the chiral and helical intralayer pairing.
\begin{figure}[h]
\subfigure{\includegraphics[width=9cm]{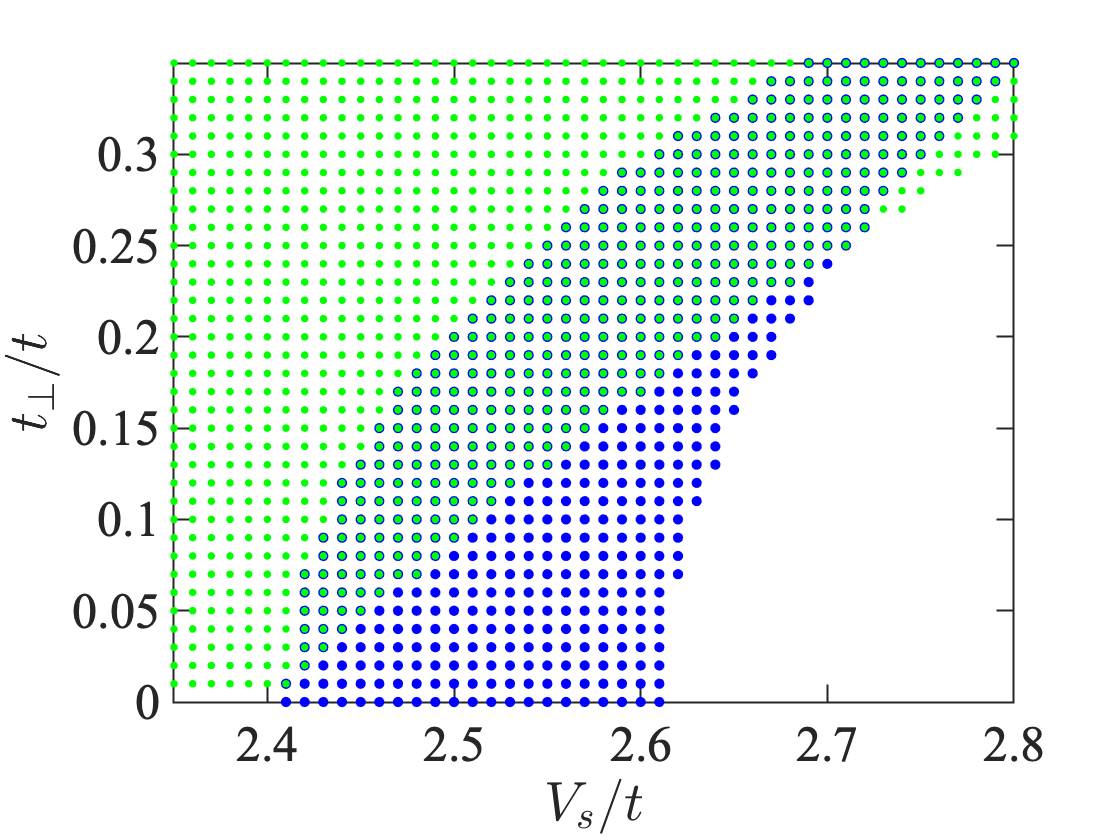}}
\caption{(color online) Determination of the phase diagram in the presence of the interlayer $s$-wave pairing interaction (see text for description).}
\label{fig:esave}
\end{figure}

Once the gap functions are obtained, the condensation energy can be calculated as
\begin{align}
\delta E = -\frac{1}{2} \sum_{\bs k} [E_{\bs k, +} + E_{\bs k, -} ] + \frac{1}{2}\sum_{ij {\bs k}} \Xi^*_{ij}(\bs k)\left [ u_{i+}(\bs k)v_{j+}^*(\bs k) + u_{i-}(\bs k)v_{j-}^*(\bs k)\right ].
\end{align}
The phase diagrams in the main text are obtained by comparing the condensation energies of the same-chirality and opposite-chirality solutions, as the latter does not necessarily have a lower energy. Take the interlayer $s$-wave pairing for example, we first obtain the region of the coexistence of the opposite-chirality $p$-wave pairing and the $s$-wave pairing, shown in Fig.~\ref{fig:esave} as the area covered by blue dots. We then plot the region where the same-chirality solutions have the lowest energy, shown Fig.~\ref{fig:esave} as the area covered by green dots. We see that this region overlaps with the coexistence region, which means the ground states of the overlapping area are in fact pure same-chirality $p$-wave pairings. Thus these two overlapping regions carve out the phase diagram shown in main text. The phase diagram for the interlayer $d$-wave pairing is determined similarly.

\section{II. Ginzburg-Landau theory}
Adopting the spinor basis $(c_{\bk 1},c_{\bk 2})^T$, the pairing function of interest takes the form,
\begin{equation}
\hat{\Xi} (\bk)=\begin{pmatrix}
\Delta_{11} f_{1\bs k} & \Delta_{12}g_{\bs k} \\
-\Delta_{12}g_{\bs k} & \Delta_{22} f_{2\bs k}
\end{pmatrix}
\label{eq:Delta}
\end{equation}
We perform a standard free energy expansion around $T_c$ in powers of the order parameter fields $\Delta_{11}$, $\Delta_{22}$ and $\Delta_{12}$. In principle, this expansion is strictly valid only when the $T_c$ of the intra- and interlayer pairings coincide. However, the qualitative picture so-obtained applies to more general scenarios. For now we take $t_\perp=0$ for simplicity. The Gorkov Greens function reads,
\begin{equation}
\hat{G}^{-1}(i w_n, \bs k) = \begin{pmatrix}
(iw_n -\xi_{\bs k}) s_0 & \hat{\Xi}_{\bs k} \\
\hat{\Xi}_{\bs k}^\dagger &(iw_n +\xi_{-\bk})s_0
\end{pmatrix} \,,
\end{equation}
where $\xi_{-\bk}= \xi_{\bk}$, $s_0$ is the rank-2 identity matrix, and $w_n=(2n+1)\pi T$ the Matsubara frequency. Defining $g_+(iw_n,\bs k) = (iw_n-\xi_{\bs k})^{-1}$, $g_-(iw_n,\bs k) = (iw_n+\xi_{\bs k})^{-1}$,  $\hat{G}_0^{-1} = \begin{pmatrix} g_+^{-1} s_0 & 0 \\ 0 & g_-^{-1} s_0  \end{pmatrix}$ and $\hat{\Xi} = \begin{pmatrix} 0 & \hat{\Xi}_{\bs k} \\ \hat{\Xi}_{\bs k}^\dagger & 0  \end{pmatrix}$, the part of the expansion essential to our discussion follows as,
\begin{eqnarray}
-\frac{T}{A} \sum_{w_n,\bk} \text{tr~ln}\hat{G}^{-1}
&=& \text{const.} + \frac{T}{2A}\sum_{w_n,\bk}\sum_{l=1}^{\infty} \frac{1}{l}\text{tr} [\hat{G}_0\hat{\Xi}]^{2l}, \nonumber
\label{eq:Expansion1}
\end{eqnarray}
where $A$ is the area of the system.
The $l=1$ term yields,
\begin{align}
 \frac{T}{2A}\sum_{w_n,\bk} \text{tr}\left[ \hat{G}_0 \hat{\Xi} \hat{G_0} \hat{\Xi}  \right]   &= \frac{T}{2A}\sum_{w_n,\bs k} g_+g_- \text{tr}\left[ \hat{\Xi}_{\bs k} \hat{\Xi}_{\bs k}^\dagger \right] \nonumber  \\
&= \frac{T}{A}\sum_{w_n,\bk} \frac{1}{w_n^2+\xi_{\bs k}^2} (|f_{1\bk}|^2|\Delta_{11}|^2+|f_{2\bk}|^2|\Delta_{22}|^2 \nonumber + 2|g_{\bk}|^2|\Delta_{12}|^2) \,.
\label{eq:Expansion2}
\end{align}
Note that the fields do not couple with one another at this level, i.e. the corresponding quadratic action takes the form $f_2=\alpha(|\Delta_{11}|^2+|\Delta_{22}|^2)+ \alpha^\prime |\Delta_{12}|^2$, where $\alpha$ and $\alpha^\prime$ can be read off from the above expression. Following the same prescription except with finite interlayer mixing $t_\perp s_1$, a coupling between $\Delta_{11}$ and $\Delta_{22}$ is induced, i.e. $\lambda \Delta^\ast_{11}\Delta_{22}+c.c.$, but only for the same-chirality state. This follows from the conservation of Cooper pair orbital angular momentum in the effective Josephson tunneling, but can also be seen from the expression,
\begin{equation}
\lambda= \frac{T}{A}\sum_{w_n,\bk} \frac{-t^2_\perp f_{1\bk}^\ast f_{2\bk}}{\left[(iw_n-\xi_{\bs k})^2-t^2_\perp \right]\left[(iw_n+\xi_{\bs k})^2-t^2_\perp \right]} \,,
\end{equation}
which vanishes upon $\bk$-summation if $f_{1\bk}$ and $f_{2\bk}$ are of opposite chirality.

Turning to the quartic order and still taking $t_\perp=0$, it is straightforward to obtain the following,
\begin{eqnarray}
f_4 &=& \beta (|\Delta_{11}|^4 + |\Delta_{22}|^4 + \alpha |\Delta_{12}|^4)  + \bar\beta(|\Delta_{11}|^2 + |\Delta_{22}|^2)|\Delta_{12}|^2  + \beta^\prime (\Delta^\ast_{11}\Delta^\ast_{22}\Delta^2_{12} + c.c.),
\end{eqnarray}
where again the coefficients can be read off from the expansion. Of particular importance to our analysis regarding the relative chirality between the $p$-wave pairings within the two layers,
\begin{equation}
\beta^\prime = -\frac{T}{2A}\sum_{w_n,\bk}\frac{f^\ast_{1\bk} f^\ast_{2\bk} g_{\bk}^2}{(w_n^2+\xi_{\bk}^2)^2}  \propto -\langle f^\ast_{1\bk} f^\ast_{2\bk} g_{\bk}^2 \rangle_\text{FS} \,.
\end{equation}
where $\langle ... \rangle_\text{FS}$ denotes a line integration across the Fermi surface. By inspection, $\beta^\prime$ is finite only when the two layers develop opposite chirality, e.g. $f_{1\bk}=\sin k_x + i \sin k_y$ and $f_{2\bk}=\sin k_x - i \sin k_y$, and in this case $\beta^\prime<0$. Therefore, in the mixed-parity phase where $\Delta_{11(22)}$ and $\Delta_{12}$ coexist, the system shall more favorably stabilize the helical $p$-wave pairing, along with appropriate relative phases between the order parameters. In this case, the system in fact possesses an exact $U(1)\times U(1)$ symmetry, where the second $U(1)$ corresponds to the relative phases between the order parameters. That is, any arbitrary gauge transformation $(\Delta_{11},\Delta_{22},\Delta_{12}) \ra (e^{2i\theta}\Delta_{11},\Delta_{22},e^{i\theta}\Delta_{12})$ leaves Eq. (3) in the main text invariant. This symmetry is retained up to all higher order terms, because any phase-dependent coupling must appear with the elementary unit of $\left( \Delta^\ast_{11}\Delta^\ast_{22}\Delta_{12}^2 + c.c.  \right)$ in order to conserve the orbital angular momentum of the Cooper pairs.

\section{III. Edge theory}

Let us start with the continuum Hamiltonian
\begin{eqnarray}
H(\bk)&=&[t(k_{x}^{2}+k_{y}^{2})-m_{0}]\tau_{z}+t_{\perp}\tau_{z}s_{x}+\Delta_{p}(k_{x}\tau_{x}-k_{y}\tau_{y}s_{z})
+[(\Delta_{d}/2)(k_{x}^{2}-k_{y}^{2})-\Delta_{s}]\tau_{y}s_{y},
\end{eqnarray}
where $\tau_{i}$ and $s_{i}$ ($i=x,y,z$) are Pauli matrices acting on the particle-hole
space and the bilayer space, respectively. Below we consider that
$t$, $m_{0}$, $t_{\perp}$, $\Delta_{p}$, $\Delta_{d}$ and $\Delta_{s}$ are all positive
for the convenience of discussion.

\subsection{Open boundary condition in the $x$ direction}

Let us first consider that the sample occupies the whole $x\geq0$ region (edge I).
As the translation symmetry is broken in the $x$ direction,
the Hamiltonian becomes
\begin{eqnarray}
H(-i\partial_{x},k_{y})=(tk_{y}^{2}-t\partial_{x}^{2}-m_{0})\tau_{z}+t_{\perp}\tau_{z}s_{x}
+\Delta_{p}(-i\partial_{x}\tau_{x}-k_{y}\tau_{y}s_{z})
+[(\Delta_{d}/2)(-\partial_{x}^{2}-k_{y}^{2})-\Delta_{s}]\tau_{y}s_{y}.
\end{eqnarray}
To simplify the analysis, we neglect the second-order $k_{y}^{2}$ term and
decompose the Hamiltonian into two parts, $H=H_{0}+H_{p}$, with
\begin{eqnarray}
H_{0}(-i\partial_{x})&=&(-t\partial_{x}^{2}-m_{0})\tau_{z}-i\Delta_{p}\partial_{x}\tau_{x},\nonumber\\
H_{p}(-i\partial_{x},k_{y})&=&-\Delta_{p}k_{y}\tau_{y}s_{z}-[(\Delta_{d}/2)\partial_{x}^{2}+\Delta_{s}]\tau_{y}s_{y}+t_{\perp}\tau_{z}s_{x}.
\end{eqnarray}
The part $H_{p}$ will be treated as a perturbation, this procedure is justified when $\Delta_{d}$, $\Delta_{s}$ and $t_{\perp}$ are
taken to be much smaller than other parameters.

Solving the eigenvalue equation
\begin{eqnarray}
H_{p}(-i\partial_{x})\Psi_{\alpha}(x)=E\Psi_{\alpha}(x)
\end{eqnarray}
with the boundary condition $\Psi(0)=\Psi(\infty)=0$, we find
that there exist two wave functions which give $E=0$, whose forms are
\begin{eqnarray}
\Psi_{1}(x)&=&\mathcal{N}e^{ik_{y}y}\sin\kappa_{1}x e^{-\kappa_{2} x}\chi_{1},\nonumber\\
\Psi_{2}(x)&=&\mathcal{N}e^{ik_{y}y}\sin\kappa_{1}x e^{-\kappa_{2} x}\chi_{2},
\end{eqnarray}
the parameters are
\begin{eqnarray}
\kappa_{1}&=&\sqrt{\frac{m_{0}}{t}-\frac{\Delta_{p}^{2}}{4t^{2}}},\quad \kappa=\Delta_{p}/2t.
\end{eqnarray}
The four-component spinors $\chi_{1,2}$ take the form
\begin{eqnarray}
\chi_{1}&=&|s_{z}=1,\tau_{y}=-1\rangle=(1,0,-i,0)^{T}/\sqrt{2},\nonumber\\
\chi_{2}&=&|s_{z}=-1,\tau_{y}=-1\rangle=(0,-1,0,i)^{T}/\sqrt{2},
\end{eqnarray}
and the normalization constant $\mathcal{N}$ takes the form
\begin{eqnarray}
\mathcal{N}&=&2\sqrt{\kappa_{2}(\kappa_{1}^{2}+\kappa_{2}^{2})/\kappa_{1}^{2}}.
\end{eqnarray}
Using perturbation theory, we find that the effective Hamiltonian describing the low-energy physics
on the edge is given by
\begin{eqnarray}
H_{\rm I}(k_{y})&=&\int_{0}^{\infty}\left(
                            \begin{array}{cc}
                              \Psi_{1}^{\dag}H_{p}\Psi_{1} & \Psi_{1}^{\dag}H_{p}\Psi_{2} \\
                              \Psi_{2}^{\dag}H_{p}\Psi_{1} & \Psi_{2}^{\dag}H_{p}\Psi_{2} \\
                            \end{array}
                          \right)dx\nonumber\\
&=&v_{y}k_{y}s_{z}+M_{\rm I}s_{y}.
\end{eqnarray}
where $v_{y}=\Delta_{p}$ and $M_{\rm I}=\Delta_{d}m_{0}/t-\Delta_{s}$. It is interesting to
find that within the first-order perturbation theory, $t_{\perp}$ does not enter
the effective Hamiltonian on the edge I.

If we consider that the sample occupies the whole $x\leq0$ region (edge III), the only difference in the wave
functions is a change of four-component spinors, i.e., $\chi_{1(2)}\rightarrow\tilde{\chi}_{1(2)}$,
with
\begin{eqnarray}
\tilde{\chi}_{1}&=&|s_{z}=1,\tau_{y}=1\rangle=(1,0,i,0)^{T}/\sqrt{2},\nonumber\\
\tilde{\chi}_{2}&=&|s_{z}=-1,\tau_{y}=1\rangle=(0,1,0,i)^{T}/\sqrt{2}.
\end{eqnarray}

Similar calculation reveals
\begin{eqnarray}
H_{\rm III}(k_{y})=-v_{y}k_{y}s_{z}+M_{\rm III}s_{y}
\end{eqnarray}
with $M_{\rm III}=M_{\rm I}$.

\subsection{Open boundary condition in the $y$ direction}

Similarly, when open boundary condition is taken in the $y$ direction, the corresponding real-space Hamiltonian is
\begin{eqnarray}
H_{0}(-i\partial_{y})&=&(-t\partial_{y}^{2}-m_{0})\tau_{z}+i\Delta_{p}\partial_{x}\tau_{y}s_{z},\nonumber\\
H_{p}(-i\partial_{y},k_{x})&=&\Delta_{p}k_{x}\tau_{x}+[(\Delta_{d}/2)\partial_{y}^{2}-\Delta_{s}]\tau_{y}s_{y}
+t_{\perp}\tau_{z}s_{x}.
\end{eqnarray}

Let us consider that the sample occupies the whole $y\geq0$ region (edge II) and solve the eigenvalue equation
\begin{eqnarray}
H_{0}(-i\partial_{y})\Psi_{\alpha}(y)=E\Psi_{\alpha}(y)
\end{eqnarray}
under the boundary condition $\Psi(0)=\Psi(+\infty)=0$, we find
there are also two wave functions which give $E=0$. Their explicit expressions
are
\begin{eqnarray}
\Psi_{1}(y)&=&\mathcal{N}e^{ik_{x}x}\sin \kappa_{1}y e^{-\kappa_{2}y}\xi_{1},\nonumber\\
\Psi_{2}(y)&=&\mathcal{N}e^{ik_{x}x}\sin \kappa_{1}y e^{-\kappa_{2}y}\xi_{2},
\end{eqnarray}
where
\begin{eqnarray}
\xi_{1}&=&|s_{z}=1,\tau_{x}=-1\rangle=(1,0,-1,0)^{T}/\sqrt{2},\nonumber\\
\xi_{2}&=&|s_{z}=-1,\tau_{x}=1\rangle=(0,i,0,i)^{T}/\sqrt{2}.
\end{eqnarray}

Then according to perturbation theory, we obtain the effective Hamiltonian on the
edge II, which is
\begin{eqnarray}
H_{\rm II}(k_{x})&=&\int_{0}^{+\infty}\left(
                                        \begin{array}{cc}
                                          \Psi_{1}^{\dag}H_{p}\Psi_{1} & \Psi_{1}^{\dag}H_{p}\Psi_{2} \\
                                          \Psi_{2}^{\dag}H_{p}\Psi_{1} & \Psi_{2}^{\dag}H_{p}\Psi_{2} \\
                                        \end{array}
                                      \right)dy\nonumber\\
&=&-v_{x}k_{x}s_{z}+M_{\rm II}s_{y}
\end{eqnarray}
where $v_{x}=\Delta_{p}$ and $M_{\rm II}=-t_{\perp}-\frac{\Delta_{d}m_{0}}{t}-\Delta_{s}$.

When the sample occupies the whole $y\leq0$ region ( edge IV), the only difference in the wave
functions is a change of four-component spinors, i.e., $\xi_{1(2)}\rightarrow\tilde{\xi}_{1(2)}$,
with
\begin{eqnarray}
\tilde{\xi}_{1}&=&|s_{z}=1,\tau_{x}=1\rangle=(1,0,1,0)^{T}/\sqrt{2},\nonumber\\
\tilde{\xi}_{2}&=&|s_{z}=-1,\tau_{x}=-1\rangle=(0,-i,0,i)^{T}\sqrt{2}.
\end{eqnarray}

Similar calculation reveals
\begin{eqnarray}
H_{\rm IV}(k_{x})&=&v_{x}k_{x}s_{z}+M_{\rm IV}s_{y}
\end{eqnarray}
where $M_{\rm IV}=t_{\perp}-\frac{\Delta_{d}m_{0}}{t}-\Delta_{s}$.

\section{IV. Mixed-parity pairing in \SRO}
We first argue the possibility of HOTSC in uniaxially strained \SRO. The $p$-wave pairing in this material can be classified into four one-dimensional representations (helical $p$-wave) and one two-dimensional $E_u$ representation \cite{Rice1995}. Accumulated evidences point to pairing in the $E_u$ channel, which forms either the widely discussed chiral $p$-wave, or a recently proposed nematic $p$-wave order \cite{Huang2018}. However, the system may be tuned towards a helical $p$-wave channel by a weak out-of-plane magnetic field \cite{Murakawa2004,Annett2008}. On the other hand, under large uniaxial strains where the system is driven towards a Lifshitz transition, the superconducting state exhibits signatures of even-parity spin-singlet pairing \cite{Hicks2014,Steppke2017}. It is hence plausible that, in the presence of the above-stated out-of-plane field, helical $p$-wave and even-parity pairings coexist at some intermediate strain strength. Noteworthily, since the helical Majorana modes in the pure helical $p$-wave phase is protected by a mirror symmetry unbroken by the out-of-plane magnetic field \cite{Ueno2013}, the 0D MZMs (therefore the topological nature) of the mixed-parity phase is insensitive to the field. As a side remark, we note a previous proposal of corner/disinclination MZMs in a pure $p$-wave model for \SRO~\cite{Benalcazar2014}.

\begin{figure}
\includegraphics[width=11.5cm]{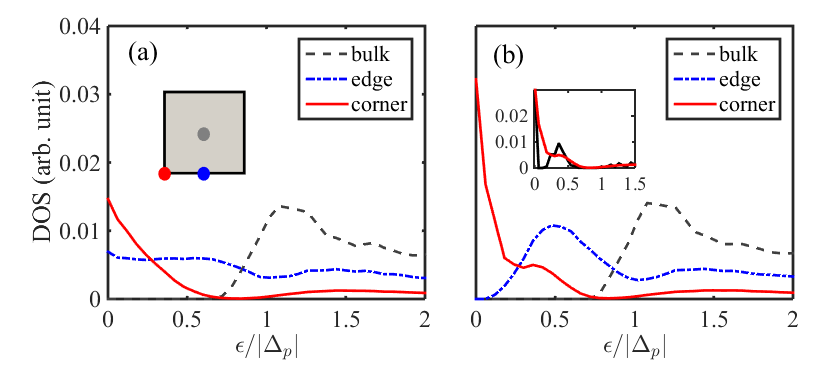}
\caption{(color online) LDOS at representative sample locations of the (a) pure helical $p$-wave and (b) mixed-parity phases. The color-coded dots in the inset of (a) indicate the locations where LDOS is measured. Calculations were performed on the same lattice as in Fig. 3 in the main text, except that we now take $\mu=-2.5t$ and $\Delta_p=0.5t$. The strain-induced even-parity pairing in (b) has $\Delta_d =0.2t$ and $\Delta_s=0.05t$. It is assumed that the mixed-parity phase occurs before the strain induces a Fermi surface Lifshitz transition in the actual experiments. The perturbation to the electronic dispersion is hence ignored for simplicity. The curves are smoothened to make the main features transparent. Note that in the unsmoothened data [black solid curve in inset of (b)], the sharp peak associated with the corner MZM [red solid curve in (b)] is detached from the secondary peak arising from gapped edge modes at $\epsilon\approx|\Delta_d|$.}
\label{fig:DOS}
\end{figure}

This topological superconducting phase can be verified by examining the local density of states (LDOS) at the edges and corners of the sample geometry. For simplicity, we consider a single-band model. The even-parity pairing shall be a mixture of $s$- and $d$-wave symmetry due to the broken $C_4$ rotational symmetry under the strain. We focus on the case with $|\Delta_d|>|\Delta_s|$, wherein corner MZMs are formed. These modes manifest as a prominent local zero-bias peak. While there could still be some ambiguity as the helical pairing shall also exhibit finite zero-bias spectrum at sample boundaries, including corners, the LDOS at the edges provide additional distinguishing features. In particular, since the even-parity pairing gaps out the helical edge modes, the mixed-parity state must observe a depletion of edge LDOS around zero-bias. These are illustrated in Fig. \ref{fig:DOS}. The opposite case with $|\Delta_d|<|\Delta_s|$ can be analyzed analogously, except that the MZMs now arise at the terminations of superconducting domains. Alternatively, the even-parity pairing may develop first at the boundaries at one strain, and then in the bulk at another larger strain. A similar sequence of transitions was noted in another context \cite{Li2017}. Finally, considering that \SRO~is a layer material, we briefly discuss about the effect of interlayer coupling. Two distinct situations are worth mentioning. One is when the interlayer coupling involves only a weak interlayer hopping, in which case the system in the mixed-parity phase is just a stack of 2D second-order topological superconductor with the Majorana modes remaining at zero energy. The other is when a weak interlayer pairing is present. In this case, the Majorana modes in general become weakly dispersive, thereby broadening the spectrum around zero bias.

In fact, there is another intriguing scenario when the bilayer spin-polarized model, Eq. (1) in the main text, is straightforwardly extended to a spinful one. All discussions on the bilayer model carry through, except that the number of MZMs doubles due to spin degeneracy. In the case of infinite layers, this model corresponds to a state with alternating chirality on each other layer. This suggests a novel mechanism to realizing a state with vanishing net surface current, in a layered material which otherwise possesses a dominant intralayer chiral $p$-wave pairing. Notably, a recent study \cite{Huang2018} pointed out a novel alternative mechanism to stabilize a time-reversal invariant nematic $p$-wave phase, which arises in the presence of a symmetry-imposed interlayer pairing.


\end{document}